\documentclass[journal]{IEEEtran}
\ifCLASSINFOpdf
  \usepackage[pdftex]{graphicx}
\else
\fi
\usepackage{amsmath}
\usepackage{amssymb}
\usepackage{xcolor, color, url, cite}
\usepackage{listings, multirow, soul}

\definecolor{dkgreen}{rgb}{0,0.6,0}
\definecolor{gray}{rgb}{0.5,0.5,0.5}
\definecolor{mauve}{rgb}{0.58,0,0.82}

\newcommand{\argmax}{\operatornamewithlimits{argmax}}

\begin{document}
%
\title{Feature selection in high-dimensional dataset using MapReduce}
%
%
%


\author{Claudio Reggiani,
        Yann-A\"el Le Borgne,
        and~Gianluca Bontempi,~\IEEEmembership{Senior Member,~IEEE}
\thanks{C. Reggiani, Y. Le Borgne, G. Bontempi are with the Machine Learning Group,
Faculty of Science, Universit\'e Libre de Bruxelles, Boulevard du Triomphe, CP 212,
1050 Brussels, Belgium
e-mail: {creggian,yleborgn,gbonte}@ulb.ac.be}}

\maketitle

\begin{abstract}
This paper describes a distributed MapReduce implementation of the minimum Redundancy Maximum Relevance algorithm, a popular feature selection method in bioinformatics and network inference problems. The proposed approach handles both \emph{tall/narrow} and \emph{wide/short} datasets. We further provide an open source implementation based on Hadoop/Spark, and illustrate its scalability on datasets involving millions of observations or features.

\end{abstract}



%
\IEEEpeerreviewmaketitle

\section{Introduction}

The exponential growth of data generation, measurements and collection in scientific and engineering disciplines leads to the availability of huge and high-dimensional datasets, in domains as varied as text mining, social network, astronomy or bioinformatics to name a few. The only viable path to the analysis of such datasets is to rely on data-intensive distributed computing frameworks \cite{bolon2015recent}.

MapReduce has in the last decade established itself as a reference programming model for distributed computing. The model is articulated around two main classes of functions, \emph{mappers} and \emph{reducers}, which greatly decrease the complexity of a distributed program while allowing to express a wide range of computing tasks. MapReduce was popularised by Google research in 2008 \cite{dean2008mapreduce}, and may be executed on parallel computing platforms ranging from specialised hardware units such as parallel field programmable gate arrays (FPGAs) and graphics processing units, to large clusters of commodity machine using for example the Hadoop or Spark frameworks \cite{dean2008mapreduce,yeung2008map,thusoo2009hive}. 

In particular, the expressiveness of the MapReduce programming model has led to the design of advanced distributed data processing libraries for machine learning and data mining, such as Hadoop Mahout and Spark MLlib. Many of the standard supervised and unsupervised learning techniques (linear and logistic regression, naive Bayes, SVM, random forest, PCA) are now available from these libraries \cite{chu2007map,mahout2012scalable,meng2016mllib}.

Little attention has however yet been given to feature selection algorithms (FSA), which form an essential component of machine learning and data mining workflows. Besides reducing a dataset size, FSA also generally allow to improve the performance of classification and regression models by selecting the most relevant features and reducing the noise in a dataset \cite{Guyon:2003:IVF:944919.944968}.


Three main classes of FSA can be distinguished: \emph{filter},  \emph{wrapper} and \emph{embedded} methods \cite{Guyon:2003:IVF:944919.944968,KohaviJohn}. Filter methods are model-agnostic, and rank features according to some metric of information conservation such as mutual information or variance. Wrapper methods use the model as a black-box to select the most relevant features. Finally, in embedded methods, feature evaluation is performed alongside the model training. In this paper, feature metrics are named hereafter \textit{feature score} functions.

Early research on distributing FSA mostly concerned wrapper methods, in which parallel processing was used to simultaneously assess different subsets of variables \cite{lopez2006solving,melab2006grid,de2006parallelizing,garcia2006parallel,conf/iwann/GuillenSMLR09}. These approaches effectively speed up the search for relevant subsets of variables, but require the dataset to be sent to each computing unit, and therefore do not scale as the dataset size increases. 

MapReduce based approaches, which address this scalability issue by splitting datasets in chunks, have more recently been proposed \cite{SFO,  Peralta2015, conf/pkdd/ZhaoCDS12,sun2014parallel, Ordozgoiti2015,Bolon-Canedo2015136,7970198}. In \cite{SFO}, an embedded approach is proposed for logistic regression. Scalability in the dataset size is obtained by relying on an approximation of the logistic regression model performance on subsets of the training set. In \cite{Peralta2015}, a wrapper method based on an evolutionary algorithm is used to drive the feature search. The first approaches based on filter methods were proposed in \cite{conf/pkdd/ZhaoCDS12, sun2014parallel}, using variance preservation and mutual information as feature selection metrics, respectively. Two other implementations of filter-based methods have lately been proposed, addressing the column subset selection problem (CSSP) \cite{Ordozgoiti2015}, and the distribution of data by features in \cite{Bolon-Canedo2015136}. Recently, a filter-based feature selection framework based on information theory \cite{Brown2012} has been implemented using Apache Spark \cite{7970198}.

In this paper we tackle the implementation of \begin{em}minimal Redundancy Maximal Relevance\end{em} (\begin{em}mRMR\end{em}) \cite{mRMR}, a forward feature selection algorithm belonging to filter methods. mRMR was shown to be particularly effective in the context of network inference problems, where relevant features have to be selected out of thousands of other noisy features \cite{bolon2015recent,journals/bmcbi/MeyerLB08}.

The main contributions of the paper are the following: \emph{i}) design of minimum Redundancy Maximum Relevance algorithm using MapReduce paradigm; \emph{ii}) customization of the feature score function during the feature selection; \emph{iii}) open-source implementation for Apache Spark available on a public repository; \emph{iv}) analysis of the scalability properties of the algorithm.

The paper is structured as follows. Section \ref{sec:mapreduce} provides an overview of the MapReduce paradigm, and Section \ref{sec:datalayout} describes the two main layouts along which data can be stored. Section \ref{sec:ifsf} presents our distributed implementation of mRMR, and details how it can be used with customised scoring functions. Section \ref{sec:results} finally provides a thorough experimental evaluation, where we illustrate the scalability of the proposed implementation by varying the number of rows and columns of the datasets, the number of selected features in the feature selection step and the number of nodes in the cluster. 

\enlargethispage{-65.1pt}

\section{MapReduce paradigm} 
\label{sec:mapreduce}



MapReduce \cite{dean2008mapreduce} is a programming paradigm designed to analyse large volumes of data in a parallel fashion. Its goal is to process data in a scalable way, and to seamlessly adapt to the available computational resources.

A MapReduce job transforms lists of input data elements into lists of output data elements. This process happens twice in a program, once for the \textit{Map} step and once for the \textit{Reduce} step. Those two steps are executed sequentially, and the Reduce step begins once the Map step is completed.

In the Map step, the data elements are provided as a list of key-value objects. Each element of that list is loaded, one at a time, into a function called \textit{mapper}. The mapper transforms the input, and outputs any number of intermediate key-value objects. The original data is not modified, and the mapper output is a list of new objects.

In the Reduce step, intermediate objects that share the same key are grouped together by a \textit{shuffling} process, and form the input to a function called \textit{reducer}. The reducer is invoked as many times as there are keys, and its value is an iterator over the related grouped intermediate values.

\begin{figure}
	\centering
	\includegraphics[width=1\linewidth]{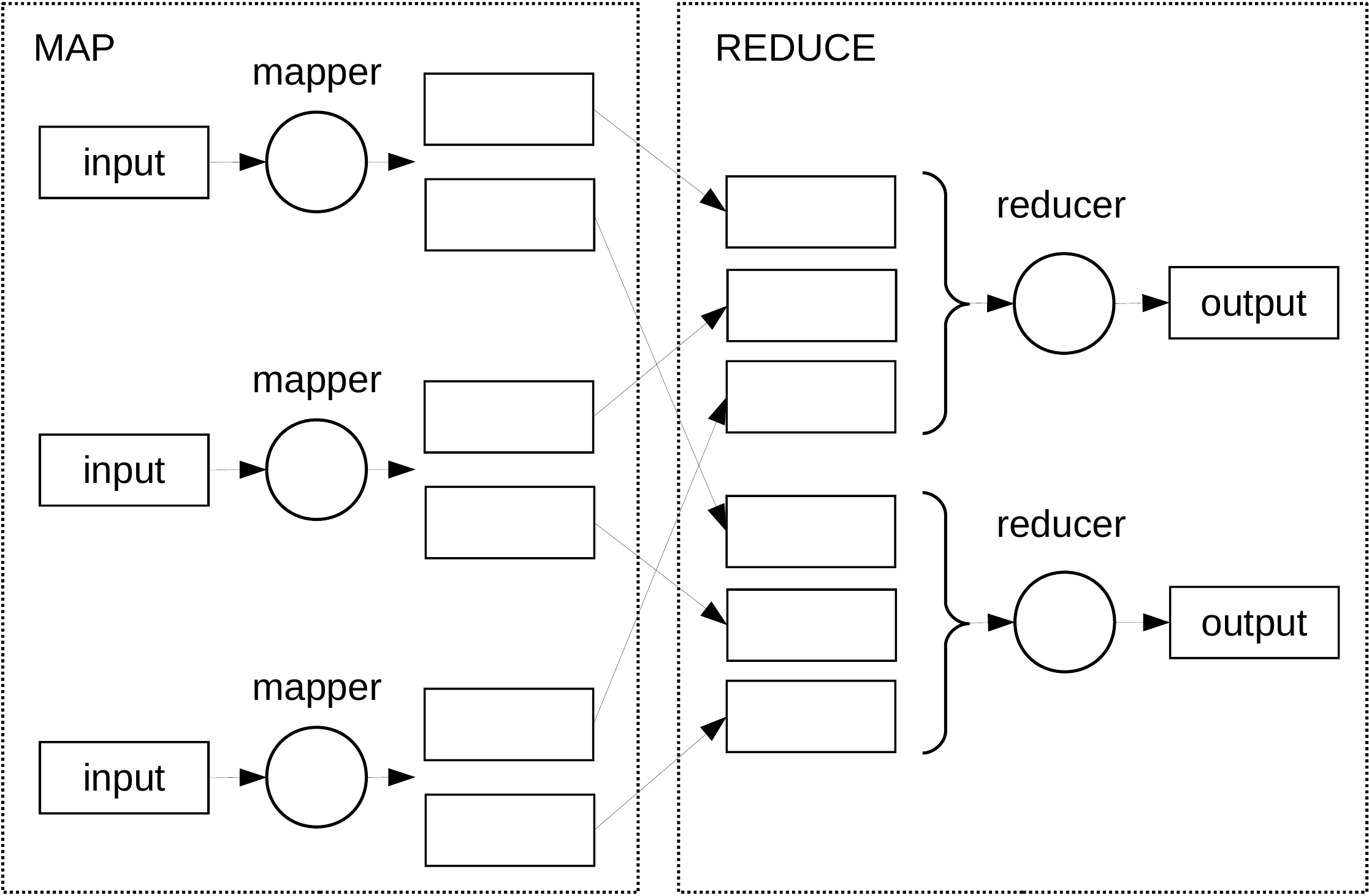}
	\caption{MapReduce overview of the data flow. The dataset stored in the distributed storage system is split into chunks across nodes (rectangular \textit{input} boxes). Each chunk is fed as input to the mapper functions, which may output intermediate objects. These objects are shuffled and grouped by keys across the network. Finally, the reducers generate the groups of intermediate objects and output the results. All objects (input, intermediate, output) are identified by a key-value pair.}
	\label{fig:mapreducesimple}
\end{figure}

Mappers and reducers run on some or all of the nodes in the cluster in an isolated environment, i.e. each function is not aware of the other ones and their task is equivalent in every node. Each mapper loads the set of files local to that machine and processes it. This design choice allows the framework to scale without any constraints about the number of nodes in the cluster. An overview of the MapReduce paradigm is reported in Figure \ref{fig:mapreducesimple}.

Algorithms written in MapReduce scale with the cluster size, and Execution Time (ET) can be decreased by increasing the number of nodes. The design of the algorithm and the data layout are important factors impacting ET \cite{Blanas:2010:CJA:1807167.1807273}.

In ET terms, jobs perform better in MapReduce when transformations are executed locally during the Map step, and when the amount of information transferred during the shuffling step is minimised \cite{sarma2013upper}. In particular, MapReduce is very well-suited for associative and commutative operators, such as \textit{sum} and \textit{multiplication}. These can indeed be partially processed using an intermediate \textit{Combine} step, which can be applied between the Map and Reduce stages.

\begin{figure}[!h]
	\centering
	\includegraphics[width=1\linewidth]{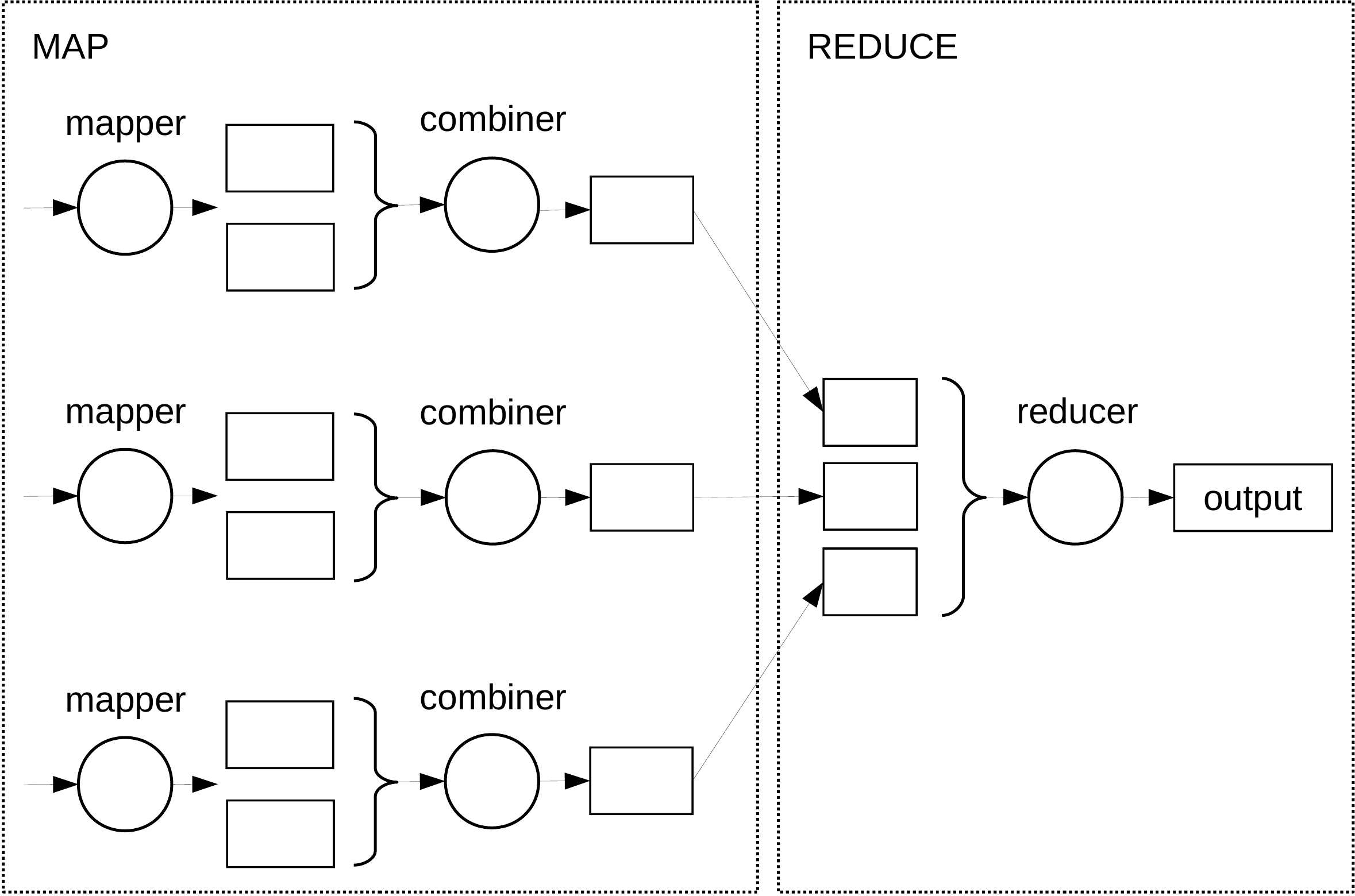}
	\caption{MapReduce overview of the data flow with the additional combiner function. Intermediate objects produced by mappers are locally aggregated by a commutative and associative function implemented in the combiner logic. In this example three, instead of six, intermediate objects are sent over the network to the reducer function. Combiners provide an efficient way to reduce the amount of shuffled data, and to reduce the overall execution time of the job.}
	\label{fig:mapreducecombiner}
\end{figure}

The combiner is an optional functionality in MapReduce \cite{dean2008mapreduce}. It locally aggregates mapper output objects before they are sent over the network. It operates by taking as input the intermediate key-value objects produced by the mappers, and output key-value pairs for the Reduce step. This optional process allows to reduce data transfer over the network, therefore reducing the global ET of the job. An illustration of the use and advantages of the combiner is given in Figure \ref{fig:mapreducecombiner}.

\section{Data Layout}
\label{sec:datalayout}

In learning problems, training data from a phenomenon is usually encoded in tables, using rows as observations, and columns as features. Let  \textit{M} be the number of observations, and \textit{N} be the number of features. Training data can be represented as a collection of vectors, $\mathbf{X}$, 
\[
\mathbf{X} = \left( \mathbf{x}_1, \mathbf{x}_2, \dots, \mathbf{x}_M \right)
\]
where
\[
\mathbf{x}_j = \left( x_{j,1}, x_{j,2}, \dots, x_{j,N} \right)\hspace*{0.5cm} \forall j \in \left(1,\dots,M\right).
\]
We will refer to this type of structure as the \textit{conventional} encoding, see Table \ref{tab:conventionalencoding}. 

It is however worth distinguishing two types of tables: \textit{tall and narrow} (T/N) tables, where $M \gg N$, and \textit{short and wide} (S/W) tables, where $M \ll N$.

The distinction is important since MapReduce divides input data in chunks of rows, that are subsequently processed by the mappers. MapReduce is therefore well suited to ingest T/N table, but much less S/W tables, since data cannot be efficiently split along columns. S/W tables are for example encountered in domains such as text mining or bioinformatics \cite{Ahn201574, JayPOHBH13}, where the number of features can be on the order of tens or hundreds of thousands, while observations may only be on the order of hundreds or thousands. 

In such cases, it can be beneficial to transform S/W into T/N tables, by storing observations as columns and features as rows. We refer to this type of structure as \textit{alternative} encoding, see Table \ref{tab:alternativeencoding}.


\begin{table}
	\centering
	\begin{tabular}{|c|c|c|c|c|} \hline
        $x_{1,1}$ & $x_{1,2}$ & \dots & \dots & $x_{1,N}$ \\ \hline
        $x_{2,1}$ & $x_{2,2}$ & \dots & \dots & $x_{2,N}$ \\ \hline
        \dots & \dots & \dots & \dots & \dots \\ \hline
        \dots & \dots & \dots & \dots & \dots \\ \hline
        $x_{M,1}$ & $x_{M,2}$ & \ldots & \dots & $x_{M,N}$ \\ \hline \multicolumn{1}{c}{}
	\end{tabular}
	\caption{Conventional encoding: Observations ($x_{i,\cdot}$) are stored along rows, and features ($x_{\cdot,j}$) are stored along columns.}
	\label{tab:conventionalencoding}
\end{table}

\begin{table}
	\centering
	\begin{tabular}{|c|c|c|c|c|} \hline
        $x_{1,1}$ & $x_{2,1}$ & \dots & \dots & $x_{M,1}$ \\ \hline
        $x_{1,2}$ & $x_{2,2}$ & \dots & \dots & $x_{M,2}$ \\ \hline
        \dots & \dots & \dots & \dots & \dots \\ \hline
        \dots & \dots & \dots & \dots & \dots \\ \hline
        $x_{1,N}$ & $x_{2,N}$ & \dots & \dots & $x_{M,N}$ \\ \hline \multicolumn{1}{c}{}
	\end{tabular}
	\caption{Alternative encoding: Observation ($x_{i,\cdot}$) are stored along columns, and features ($x_{\cdot,j}$) are stored along rows.}
	\label{tab:alternativeencoding}
\end{table}


\section{Iterative feature selection framework}
\label{sec:ifsf}




This section first recalls the standard mRMR algorithm \cite{mRMR}. We then detail our MapReduce implementation, for both conventional and alternative encodings. We conclude with an example of custom feature score implementation using the Pearson correlation coefficient.

\subsection{minimal Redundancy Maximal Relevance}

Let us define the dataset as the table $\mathbf{X}$ with $M$ rows, $N$ columns and discrete values. We define $\mathbf{x}_{k}$ as the $k-th$ column vector of the dataset and $\mathbf	{c}$ as the class vector. Furthermore, let us define $L$ as the number of features to select and $i_{c}^{l}$ and $i_{s}^{l}$ as the sets at step $l$ ($1 \leqslant l \leqslant L$) of candidate and selected features indices, respectively. At $l=1$, we have $i_{c}^{1} = \left\{ 1,...,N \right\}$ and $i_{s}^{1} = \varnothing$. The pseudo-code of the algorithm is reported in Listing \ref{lst:mrmr}.

\begin{lstlisting}[caption={minimum Redundancy Maximum Relevance Pseudo-code. $I(\cdot)$ is the function that, given two vectors, returns their mutual information. $\mathbf{x}_{k}$ is the $k-th$ column vector of the dataset and $\mathbf{c}$ is the class vector. $L$ is the number of features to select, $i_{c}^{l}$ and $i_{s}^{l}$ as the sets at step $l$ ($1 \leqslant l \leqslant L$) of candidate and selected features indices.\\},label={lst:mrmr},mathescape,language=Java]
$i_{c}^{1} = \left\{1,...,N \right\}$
$i_{s}^{1} = \varnothing$
for $l=1 \to L$
  for $k \in i_{c}^{l}$
    $I_{\mathbf{x}_{k}, \mathbf{c}} \leftarrow I\left(\mathbf{x}_{k}, \mathbf{c}\right)$
    for $j \in i_{s}^{l}$
      $I_{\mathbf{x}_{k}, \mathbf{x}_{j}} \leftarrow I\left(\mathbf{x}_{k}, \mathbf{x}_{j}\right)$
    $g_k$ $\leftarrow$ $I_{\mathbf{x}_{k}, \mathbf{c}} - \frac{1}{\left|i_{s}^{l}\right|} \sum_{j \in i_{s}^{l}} I_{\mathbf{x}_{k}, \mathbf{x}_{j}}$
  $k^{*}$ $\leftarrow$ argmax($g_k$)
  $i_{c}^{l+1} \leftarrow i_{c}^{l} \backslash k^{*}$
  $i_{s}^{l+1} \leftarrow i_{s}^{l} \cup k^{*}$
output $i_{s}^{L}$
\end{lstlisting}

mRMR is an iterative greedy algorithm: at each step the candidate feature is selected based on a combination of its mutual information with the class and the selected features:
\begin{equation}
\label{eq:scoremrmr}
\begin{array}{c}
\argmax_{k\in i_{c}^{l}}g_k\left(\cdot\right)\\\\

g_k\left(\cdot\right)=\begin{cases}
I\left(\mathbf{x}_{k};\mathbf{c}\right) & l = 1 \\
I\left(\mathbf{x}_{k};\mathbf{c}\right) - \frac{1}{\left|i_{s}^{l}\right|} \sum_{j \in i_{s}^{l}} I\left(\mathbf{x}_{k};\mathbf{x}_{j}\right) & l > 1
\end{cases}\\

\end{array}
\end{equation}

where $I(\cdot)$ returns the mutual information of two vectors. The \textit{feature score} $g\left(\cdot\right)$ is assessed in Lines 5-8 in Listing \ref{lst:mrmr}.

We redesigned the algorithm using MapReduce paradigm on Apache Spark, distributing the feature evaluation into the cluster. Besides the core features of MapReduce previously described, our design takes advantage of the \textit{broadcast} operator provided in Apache Spark. Broadcasted variables are commonly used in machine learning algorithms to efficiently send additional data to every mapper and reducer as read-only variables \cite{Karau:2015:LSL:2717070}.


\subsection{mRMR in MapReduce with conventional encoding}

Let us define the dataset as a Resilient Distributed Dataset (RDD) \cite{DBLP:conf/nsdi/2012} of $M$ tuples $\left(\mathbf{x}, c\right)$, where $\mathbf{x}$ is the input (observation) vector and $c$ is the target class value.

Considering the dataset with only discrete values, we represent with $d_c$ the set of categorical values of the class, and with $d_v$ the (union) set of unique categorical values of all features. If the dataset has binary values, then $d_c = d_v = \left\{ 0,1 \right\}$. In case of having features with different sets of categorical values, then $d_v$ is the union of unique categorical values of all features.

\begin{figure}[!ht]
	\centering
	\includegraphics[width=0.75\linewidth]{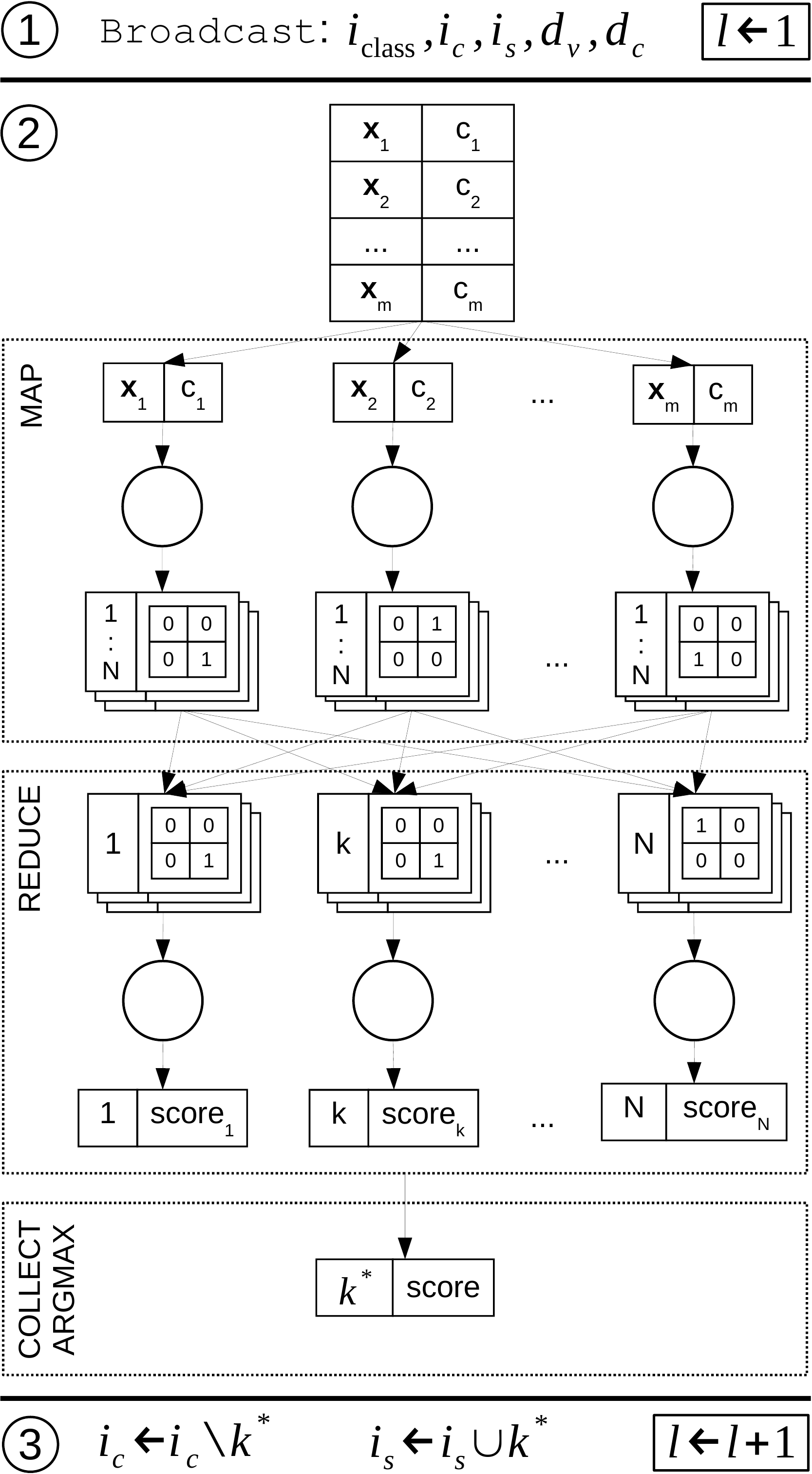}
	\caption{Illustrative representation of the first iteration of a MapReduce job with discrete values using the conventional encoding. There are as many iterations as the number of features to select. At each iteration, each mapper outputs $N-l+1$ contingency tables for every combination of candidate features and class vector, and, from the second iteration, $(N-l+1)*\left|i^l_s\right|$ contingency tables for every combination of candidate and selected features.}
	\label{fig:TraditionalEncodingSchema}
\end{figure}

The input vector is partitioned in candidate and selected features, labeled respectively as  $\mathbf{x}_{c}$ and $\mathbf{x}_{s}$ ($\mathbf{x}_{c} \cup \mathbf{x}_{s} = \mathbf{x}$, $\left|\mathbf{x}\right|=N$). Variables $L$, $i_{c}^{l}$ and $i_{s}^{l}$ are defined as in the previous section and $i_{class}$ is the class column index. Listings \ref{lst:mrmrcolwise}, \ref{lst:mrmrcolwisemap} and \ref{lst:mrmrcolwisereduce} report the MapReduce job, the mapper and reducer functions, respectively, while an illustrative overview of the data flow is reported in Fig. \ref{fig:TraditionalEncodingSchema}.

\begin{lstlisting}[caption={mRMR MapReduce job with conventional data encoding. $L$ is the number of features to select, $i_{c}^{l}$ and $i_{s}^{l}$ are the sets at step $l$ ($1 \leqslant l \leqslant L$) of candidate and selected features indices. $i_{class}$ is the class column index. $d_c$ is the set of categorical values of the class, and $d_v$ is the (union) set of unique categorical values of all features.\\},label={lst:mrmrcolwise},mathescape,language=Java]
$i_{c}^{1} = \left\{1,...,N \right\}$
$i_{s}^{1} = \varnothing$
for $l=1 \to L$
  broadcast $i_{class}$, $i_{c}^{l}$, $i_{s}^{l}$, $d_v$, $d_c$
  scores <- mapreduce(RDD, mapper, reducer)
  $k^{*}$ $\leftarrow$ collectArgmax(scores)
  $i_{c}^{l+1} \leftarrow i_{c}^{l} \backslash k^{*}$
  $i_{s}^{l+1} \leftarrow i_{s}^{l} \cup k^{*}$
output $i_{s}^{L}$
\end{lstlisting}
\begin{lstlisting}[caption={mRMR MapReduce mapper function with conventional data encoding. $i_{c}^{l}$ and $i_{s}^{l}$ as the sets at step $l$ ($1 \leqslant l \leqslant L$) of candidate and selected features indices. $i_{class}$ is the class column index. $d_c$ is the set of categorical values of the class, and $d_v$ is the (union) set of unique categorical values of all features. $e$ is a single observation fed as input to the mapper, $k$ and $j$ represent column indices and \textit{contTable} is the function that creates a contingency table.\\},label={lst:mrmrcolwisemap},mathescape,language=Java]
# broadcasted vars: $i_{class}$, $i_{c}^{l}$, $i_{s}^{l}$, $d_v$, $d_c$
mapper($\cdot$, $e$)
  for $k \in i_{c}^{l}$
    output ($k$, contTable($e_k$, $e_{i_{class}}$, $d_v$, $d_c$))
    for $j \in i_{s}^{l}$
      output ($k$, contTable($e_k$, $e_j$, $d_v$, $d_c$))
\end{lstlisting}
\begin{lstlisting}[caption={mRMR MapReduce reducer function with conventional data encoding. $k$ is a column index and $t$ is a collection of contingency tables. The \textit{score} function process all the contingency tables associated with the column with index $k$ and return the feature score.\\},label={lst:mrmrcolwisereduce},mathescape,language=Java]
reducer($k$, $t$)
  output($k$, score($t$))
\end{lstlisting}

For every $\left(e_k, e_{i_{class}}\right)$ pair, the mapper task outputs a contingency table, \textit{contTable}, with rows defined as the categorical values in $d_c$ and columns defined as the categorical values in $d_v$. The element corresponding to row $e_{i_{class}}$ and column $e_k$ is set to 1, while all the others are set to 0. Considering the dataset in Table \ref{tab:datatable}, having one binary class column and four categorical features (with three possible values: \textit{-2}, \textit{0}, \textit{2}), an example of emitted contingency table is reported in Table \ref{tab:conttable}. In this example the class vector can only have two possible values: \textit{0} and \textit{1}; any feature can only have three possible values: \textit{-2}, \textit{0} and \textit{2}. The input pair $\left(e_k, e_{i_{class}}\right)$ is $\left(2,0\right)$, therefore the element corresponding to row \textit{0} and column \textit{2} is set to 1, all the others are set to 0.

In case of $\left(e_k, e_j\right)$ pair, the contingency table has both rows and columns defined by categorical values in $d_v$.

\begin{table}
	\centering
	\begin{tabular}{c|c|c|c|c|c|} \cline{2-6}
		\multicolumn{1}{ c| }{\#entry} & \multicolumn{1}{ |c| }{class} & \multicolumn{4}{ c| }{features} \\ \cline{2-6}
         & c & $\mathbf{x}_1$ & $\mathbf{x}_2$ & $\mathbf{x}_3$ & $\mathbf{x}_4$ \\ \cline{2-6}
        1 & 0 & 2 & 0 & 0 & -2 \\ \cline{2-6}
        2 & 0 & 0 & -2 & 2 & 0 \\ \cline{2-6}
        3 & 0 & 0 & 2 & 0 & -2 \\ \cline{2-6}
        4 & 1 & -2 & 0 & 0 & 0 \\ \cline{2-6}
        \dots & \dots & \dots & \dots & \dots & \dots \\ \cline{2-6}
        \multicolumn{1}{c}{}
	\end{tabular}
	\caption{Example of dataset encoded with conventional layout.}
	\label{tab:datatable}
\end{table}

\begin{table}
	\centering
	\begin{tabular}{c|c|c|c|c|} \cline{3-5}
		\multicolumn{1}{ c }{} & & \multicolumn{3}{ c| }{$d_v$} \\ \cline{3-5}
		\multicolumn{1}{ c }{} & & \textbf{-2} & \textbf{0} & \textbf{2} \\ \cline{1-5}
		\multicolumn{1}{ |c  }{\multirow{2}{*}{$d_c$} } & \multicolumn{1}{ |c| }{\textbf{0}} & 0 & 0 & 1 \\ \cline{2-5}
		\multicolumn{1}{ |c  }{} 			& \multicolumn{1}{ |c| }{\textbf{1}} & 0 & 0 & 0 \\ \cline{1-5} \multicolumn{1}{c}{}
	\end{tabular}
	\caption{Contingency table emitted by the mapper function as a result of processing the pair $\left(\mathbf{x}_1,c\right)$ of the first entry in Table \ref{tab:datatable}.}
	\label{tab:conttable}
\end{table}


At the cost of managing discrete values only, the commutative and associative properties of the contingency table allows the use of the combiner function, thus minimizing the amount of data exchanged across the cluster during shuffling. While the single mapper outputs one or more contingency tables for each candidate feature, those tables emitted by mappers executed on a given node can be locally reduced via the Combine step. Assuming that the first four entries in Table \ref{tab:datatable} are processed by four mappers in the same machine, Table \ref{tab:conttable2} is the result of the combiner after the aggregation of four contingency tables of the $\mathbf{x}_1$ feature produced by the mappers. In this example, the combiner performs an element-wise sum of the contingency tables given as input.



\begin{table}
	\centering
	\begin{tabular}{c|c|c|c|c|} \cline{3-5}
		\multicolumn{1}{ c }{} & & \multicolumn{3}{ c| }{$d_v$} \\ \cline{3-5}
		\multicolumn{1}{ c }{} & & \textbf{-2} & \textbf{0} & \textbf{2} \\ \cline{1-5}
		\multicolumn{1}{ |c  }{\multirow{2}{*}{$d_c$} } & \multicolumn{1}{ |c| }{\textbf{0}} & 0 & 2 & 1 \\ \cline{2-5}
		\multicolumn{1}{ |c  }{} & \multicolumn{1}{ |c| }{\textbf{1}} & 1 & 0 & 0 \\ \cline{1-5} \multicolumn{1}{c}{}
	\end{tabular}
	\caption{Aggregated contingency table emitted by the combiner function as a result of processing the pair $\left(\mathbf{x}_1,c\right)$ of the first four entries in Table \ref{tab:datatable}.
}
	\label{tab:conttable2}
\end{table}

\subsection{mRMR in MapReduce with alternative encoding}

Data stored in alternative encoding has one column per observation and one row per feature. In this case, let us define the dataset as a RDD of $N$ tuples $\left(k, \mathbf{x}\right)$, where $\mathbf{x}$ is the feature vector and $k$ is the row index ($k \in \left\{1,...,N\right\}$). Feature and class values could be discrete and continuous as well. With respect to the design of mRMR in MapReduce with conventional encoding, a set of vectors are broadcasted across the cluster: $\mathbf{v}_{class}$ is the class vector, $v_{s}$ is the collection of selected feature vectors and $i_{s}$ is the collection of selected feature indices. Variable $L$ is defined as in the previous section and \textit{getEntry} function is a MapReduce task that retrieves the feature vector from the RDD, given a feature index. Listings \ref{lst:mrmrrowwise} and \ref{lst:mrmrrowwisemap} report the MapReduce job and the mapper function, respectively.

\begin{lstlisting}[caption={mRMR MapReduce job with alternative data encoding. \textit{RDD} represents the distributed dataset and $L$ is the number of features to select. $\mathbf{v}_{class}$ is the class vector, $v_{s}$ is the collection of selected feature vectors and $i_{s}$ is the collection of selected feature indices. The \textit{getEntry} function retrieves the feature vector from the RDD, given a feature index.\\},label={lst:mrmrrowwise},mathescape,language=Java]
$i_{s}^{1} = \varnothing$
$v_{s}^{1} = \varnothing$
for $l=1 \to L$
  broadcast $\mathbf{v}_{class}$, $i_{s}^{l}$, $v_{s}^{l}$
  scores <- mapreduce(RDD, mapper)
  $k^{*}$ $\leftarrow$ collectArgmax(scores)
  $v^{*}$ <- getEntry(RDD, $k^{*}$)
  $i_{s}^{l+1} \leftarrow i_{s}^{l} \cup k^{*}$
  $v_{s}^{l+1} \leftarrow v_{s}^{l} \cup v^{*}$
output $i_{s}^{L}$
\end{lstlisting}
\begin{lstlisting}[caption={mRMR MapReduce mapper function with alternative data encoding. $\mathbf{v}_{class}$ is the class vector, $v_{s}$ is the collection of selected feature vectors. The tuple $\left(k, \mathbf{x}\right)$ is composed by the feature vector, $\mathbf{x}$, and the feature index, $k$. The \textit{score} function processes the vectors and returns the feature score.\\},label={lst:mrmrrowwisemap},mathescape,language=Java]
# broadcasted variables: $\mathbf{v}_{class}$, $v_{s}^{l}$
mapper($\cdot$, ($k$, $\mathbf{x}$))
  score <- score($\mathbf{x}$, $\mathbf{v}_{class}$, $v_{s}^{l}$)
  output ($k$, score)
\end{lstlisting}

\begin{figure}[!ht]
	\centering
	\includegraphics[width=0.75\linewidth]{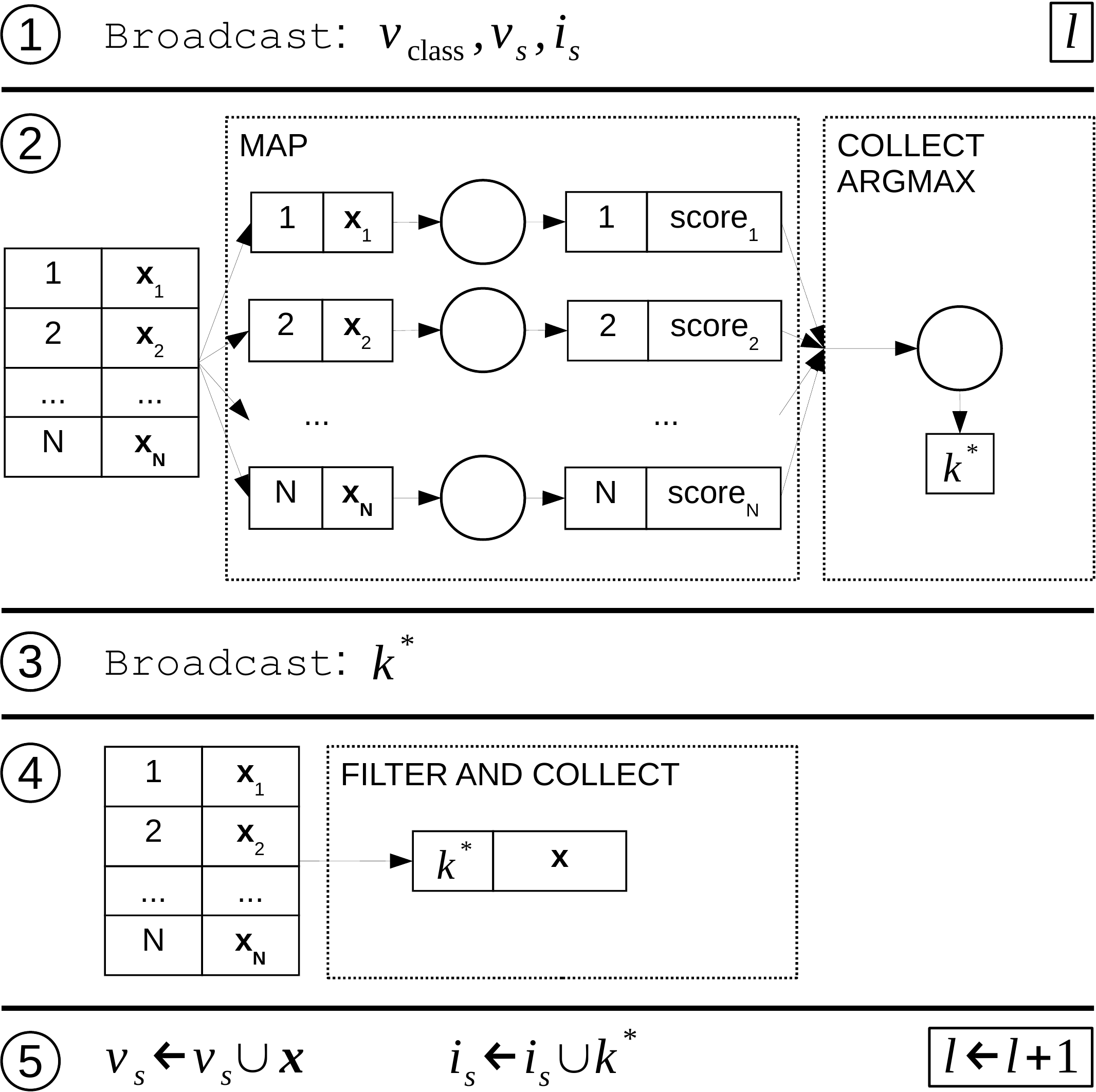}
	\caption{Illustrative representation of a single iteration of a MapReduce job with alternative encoding. Steps 1-5 represent one iteration of the loop; there are as many iterations as the number of features to select.}
	\label{fig:AlternativeEncodingSchema}
\end{figure}

While in conventional encoding we used the contingency table as intermediate data structure, the design of mRMR in MapReduce with alternative encoding broadcasts at each iteration all required data for calculation to mappers. This design provides two main advantages: it deals with both discrete and continuous features as well, and the MapReduce job is composed by the Map step only. At the small cost of broadcasting some variables, all operations are executed locally. An illustrative overview of the data flow is reported in Fig. \ref{fig:AlternativeEncodingSchema}.



\subsection{Custom score with alternative encoding}

By introducing mRMR in MapReduce with alternative encoding, we propose a solution to store and analyse high-dimensional datasets in distributed environments. While mRMR is a well-known feature selection method, some problems might require a different feature score function to perform the selection of features.
For this reason, we provide an interface to customize the feature score function, see Listing \ref{lst:interface}.
This interface is available only with alternative encoding as the conventional one is constraint by working with discrete values. 
By means of the \textit{getResult} function, the interface provides at each iteration of the algorithm three variables: the candidate feature vector \textit{variableArray}, the class vector \textit{classArray} and the collection of selected feature vectors \textit{selectedVariablesArray} (respectively $\mathbf{x}$, $\mathbf{v}_{class}$ and ${v}^{l}_{s}$ in Listing \ref{lst:mrmrrowwisemap}). The function has to return a scalar value representing the feature score for the candidate feature at such iteration.



\begin{lstlisting}[caption={Scala interface for custom feature score.\\},label={lst:interface},mathescape,language=Java]
function getResult:
  arguments:
    variableArray: Array[Double]
    classArray: Array[Double]
    selectedVariablesArray: Array[Array[Double]]
  return: Double
\end{lstlisting}

The mRMR design in MapReduce with alternative encoding described in this paper is implemented using the \textit{getResult} function. Its mathematical representation (as described in the original paper \cite{mRMR}) is reported in Formula \ref{eq:scoremrmr}, where $g\left(\cdot\right)$ defines the feature score function.

To illustrate the flexibility of the framework for feature selection with alternative encoding, we provide an example of custom feature score function implementation: an approximation of the mutual information, $f\left(\cdot\right)$, by means of the Pearson correlation coefficient (Formula \ref{eq:scorecorr} and Listing \ref{lst:correlation}).




\begin{equation}
\label{eq:scorecorr}
\begin{array}{c}
\argmax_{k\in i_{c}^{l}}g_k\left(\cdot\right)\\\\

g_k\left(\cdot\right)=\begin{cases}
f\left(\mathbf{x}_{k};\mathbf{c}\right) & l = 1 \\
f\left(\mathbf{x}_{k};\mathbf{c}\right) - \frac{1}{\left|i_{s}^{l}\right|} \sum_{j \in i_{s}^{l}} f\left(\mathbf{x}_{k};\mathbf{x}_{j}\right)  & l > 1
\end{cases}\\

\end{array}
\end{equation}

\begin{lstlisting}[caption={Pseudo-code implementation of the \textit{getResult} interface for customizing the feature score, based on Pearson correlation coefficient instead of mutual information. \textit{pcc} calculates the Pearson correlation given two input vectors. The \textit{size} function returns the number of elements in the collection.\\},label={lst:correlation},mathescape]
cor2mi(v)
  -0.5 * log( 1.0-(v*v) )

getResult(variableArray, classArray, selectedVariablesArray)
  val v <- variableArray
  val clv <- classArray
  val sva <- selectedVariablesArray
  
  sc <- cor2mi(pcc(v, clv)) # f($\cdot$)

  sfs <- 0
  for i=0 $\to$ size(sva)
    sv <- sva[i]
    sfs <- sfs + cor2mi(pcc(v, sv))

  coeff <- 1.0
  if size(sva) > 1
    coeff <- 1.0 / size(sva)

  return sc - (coeff * sfs)
\end{lstlisting}

The interface should be implemented as a third-party library, packaged as a \textit{jar} file, and provided as input during the spark submission job. The implementation of the custom score which approximates the mutual information by means of the Pearson correlation coefficient is provided as an example in the public repository (https://github.com/creggian/spark-ifs).

\section{Results}
\label{sec:results}

The source code of mRMR implementation in MapReduce with both encodings is available as a Scala library, along with examples, on a public repository (https://github.com/creggian/spark-ifs).

We studied the scalability of the implementation of mRMR in MapReduce in both encodings in a cluster with the following specifications: Hadoop cluster of 10 nodes, where each node has Dual Xeon e5 2.4Ghz processor, 24 cores, 128GB RAM and 8TB hard disk; all nodes are connected with a 1Gb ethernet connection. Using Apache Spark v1.5.0, we submit jobs with 4GB of RAM for both the driver and the executors.

For the evaluation of mRMR implementations we used binary artificial datasets. We followed the principles of CorrAL dataset \cite{bolon2015feature}, in which four features determine the class value with the following formula: $c = \left(x_1 \wedge x_2\right) \vee \left(x_3 \wedge x_4\right)$, one is irrelevant and the last one is partially correlated with the class. In all our datasets, the class value ($c$) depends on the value of 8 features (Formula \ref{eq:1}); the remainings are irrelevant.
\begin{equation}
\label{eq:1}
c = \left(\left(x_1 \wedge x_2\right) \vee \left(x_3 \wedge x_4\right)\right) \wedge \left(\left(x_5 \wedge x_6\right) \vee \left(x_7 \wedge x_8\right)\right)
\end{equation}

We assessed the scalability on the number of rows, the number of columns, the number of selected features, and the number of nodes. We used two kinds of dependent variables: the relative execution time per executor and the computational gain. The former is the ratio between ET divided by ET of \texttt{1x}, the latter is the ratio between ET of 1-node and ET. We ran the tests three times to assess the variability of the results; in all figures the maximum, minimum and mean of these three values are connected through a solid vertical line.

\subsection{Scalability across the number of rows}

We tested the scalability on the number of rows by means of four datasets, each with 1000 columns and an increasing number of rows: 1M, 4M, 7M and 10M (M = millions). We configured the cluster and the algorithm to select 10 features in a distributed environment of 10 nodes (Fig. \ref{fig:rows}).

\subsection{Scalability across the number of columns}

We assessed the scalability on the number of columns using four datasets, each with 1M rows and an increasing number of columns: 100, 400, 700 and 1000. We configured the cluster and the algorithm to select 10 features in a distributed environment of 10 nodes (Fig. \ref{fig:cols}).

\subsection{Scalability across the number of selected features}

\begin{figure}
	\centering
	\includegraphics[width=1\linewidth]{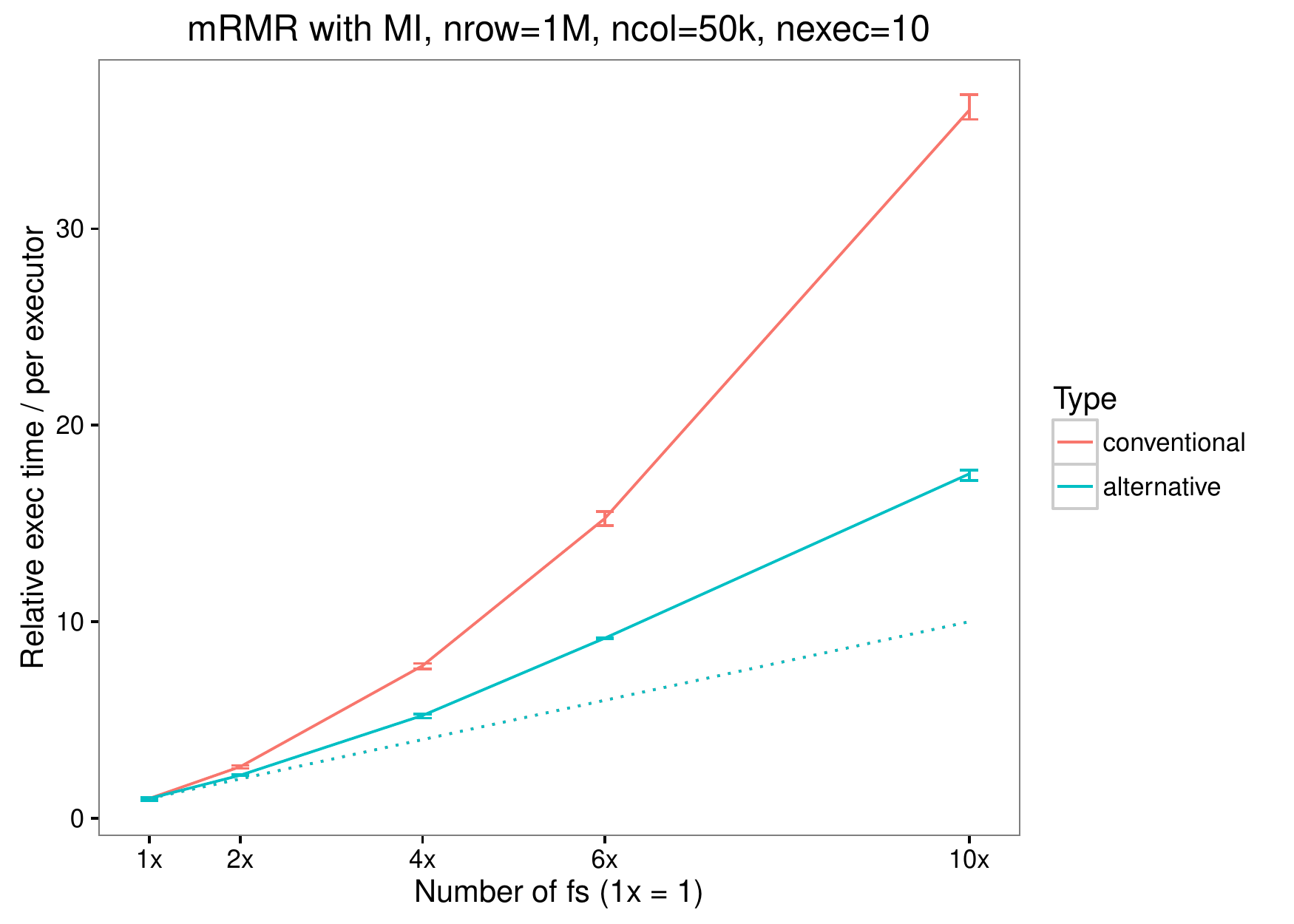}
	\caption{mRMR scalability across the number of selected features.}
	\label{fig:nfs}
\end{figure}

We investigated the scalability on the number of selected features using a dataset with 1M rows and 50k (k = thousands) columns. We parametrised the cluster to distribute the computation over 10 nodes, and the algorithm to select an increasing number of features: 1, 2, 4, 6, 10 (Fig. \ref{fig:nfs}).

\subsection{Scalability across the number of nodes}

\begin{figure}
	\centering
	\includegraphics[width=1\linewidth]{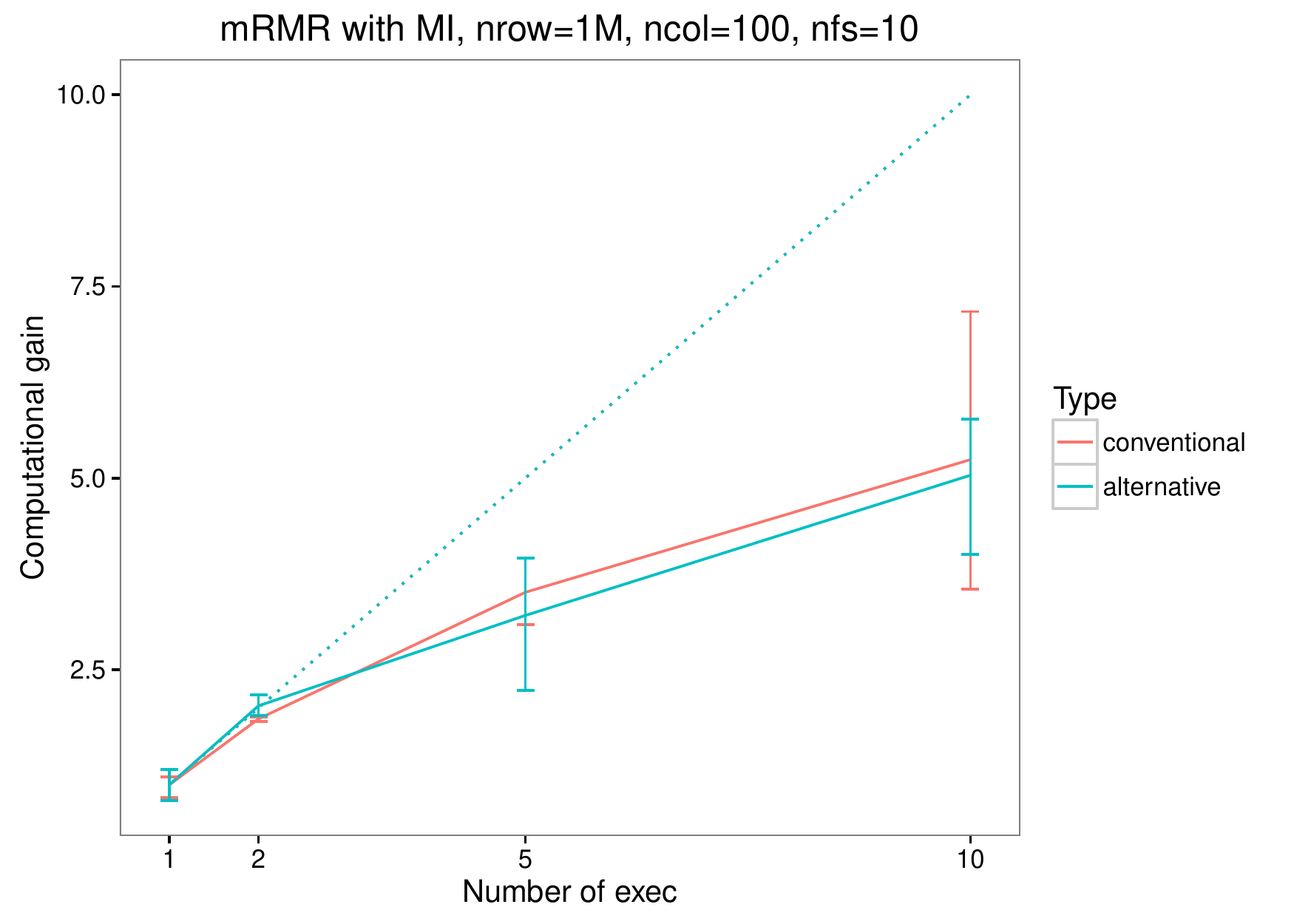}
	\caption{mRMR scalability across the number of nodes.}
	\label{fig:exec}
\end{figure}

We tested the scalability across the number of nodes using a dataset with 1M rows and 100 columns. We configure the algorithm to select 10 features, and the cluster to distribute the work over 1, 2, 5 and 10 nodes (Fig. \ref{fig:exec}).

\vspace{\baselineskip}
By comparing the linear scalability (dotted line) with the actual performances, results show that the scalability of mRMR in MapReduce is linear with respect to the number of rows, as expected by MapReduce design; superlinear with respect to the number of columns; sublinear with respect to the number of selected features and nodes, as expected by our iterative algorithm design and the increasing amount of data exchanged in the network with the increasing of nodes, respectively.

\begin{figure}
	\centering
	\includegraphics[width=1\linewidth]{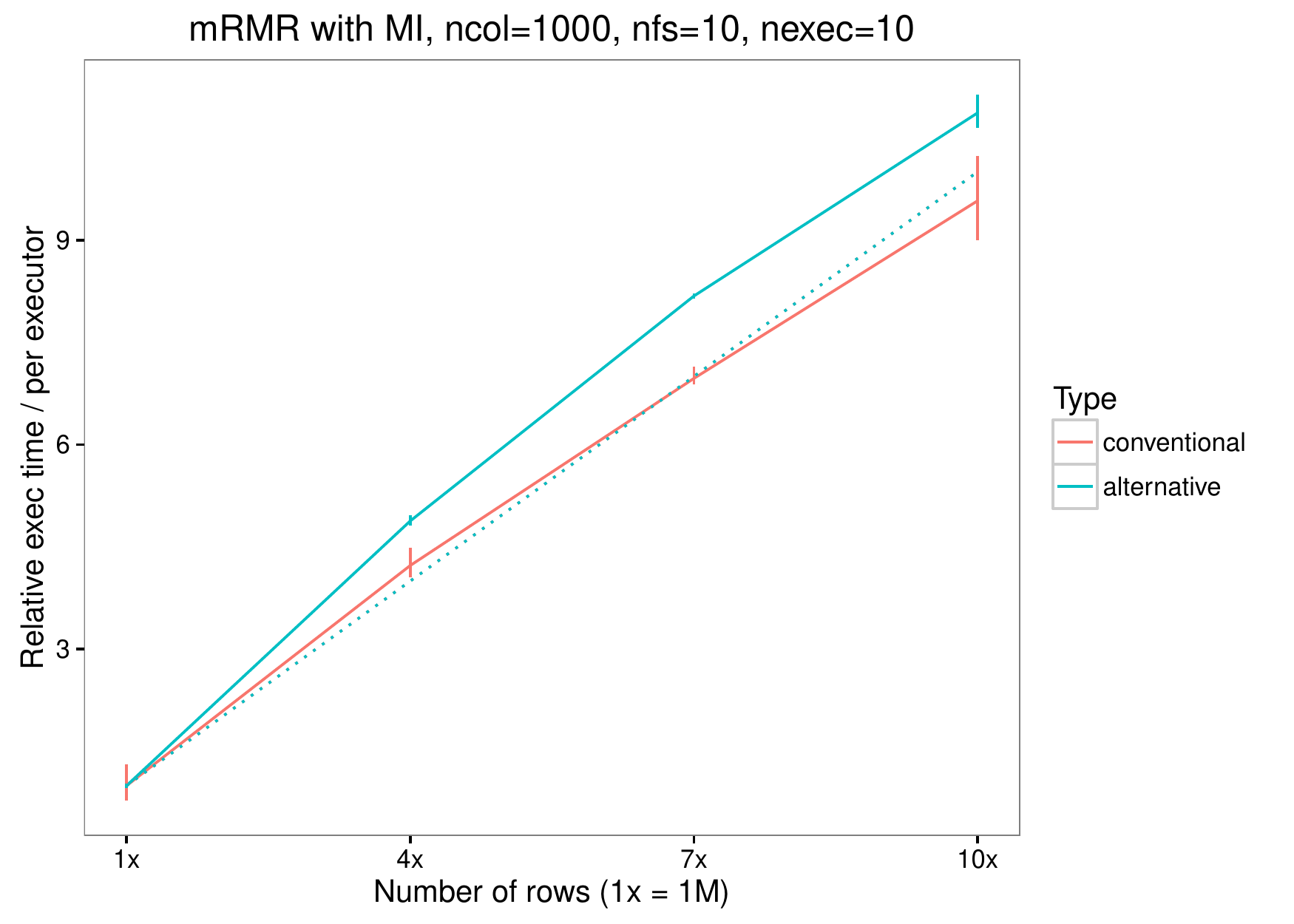}
	\caption{mRMR scalability across the number of rows.}
	\label{fig:rows}
\end{figure}

\begin{figure}
	\centering
	\includegraphics[width=1\linewidth]{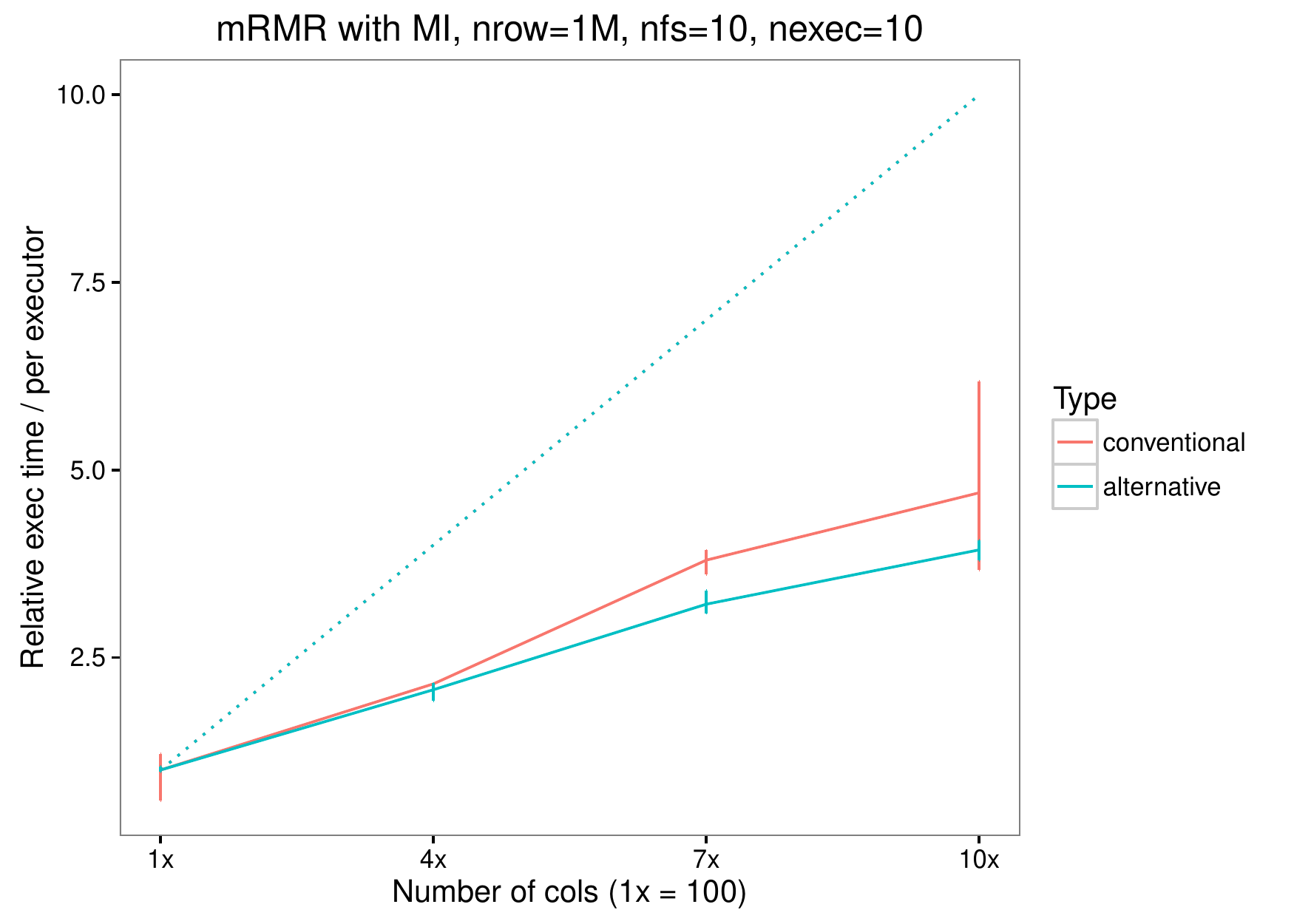}
	\caption{mRMR scalability across the number of columns.}
	\label{fig:cols}
\end{figure}

In studying mRMR with conventional and alternative layouts, we chose to use as independent variable the number of rows (columns) instead of the number of observation (features) for the following reason: while in the conventional layout we are able to scale across a very large number of rows, in the alternative layout we are strictly constraint by the amount of memory available in the mapper task to scale across the number of columns. In Figures \ref{fig:rows} and \ref{fig:cols}, we tested up to 10 million rows and up to one thousand columns, because very high-dimensional S/W tables raises memory errors in the cluster. Hence, even though we show the relative execution time, it would be incorrect to plot performances by increasing the number of observation (features).

The absolute execution time of mRMR MapReduce jobs with alternative encoding is generally 4-6x faster than the respective jobs with conventional encoding.

\section{Conclusion}
\label{sec:conclusion}

In this work we investigated the design and scalability of mRMR algorithm in MapReduce. We proposed two implementations depending on the data layout and we provide an interface in order to customize the feature score function in the alternative encoding scenario. Despite Hadoop limitations for handling data with a large number of columns, the alternative data layout is a solution to store data from a phenomenon that has a very large number of features. 
In both conventional and alternative data layouts, we studied the scalability of mRMR in different settings: the number of rows, columns, selected features and nodes. The results give an overview of the performance of the algorithm implemented in MapReduce.

Currently, the conventional encoding works only with discrete values; its extension to continuous features would require either additional MapReduce jobs to estimate the binning strategy or additional parameters as input of the jobs. MapReduce jobs with conventional encoding perform slower than those in alternative encoding; the latter is also by design inherently flexible to work with both discrete and continuous features.

The design described in this paper uses dense data structures to read the input and to produce intermediate objects. We improved the performance of the framework so to handle sparse data as well. The result is an update of the source code in the public repository, while the source code used to produce the results of this paper is available at https://github.com/creggian/spark-ifs/tree/7cb63bfc2b5d224fb830c3523687bd38bac66f97.

With respect to \cite{7970198}, we provide a different distributed implementation of the mRMR algorithm. It works with both traditional and alternative encoding, with the possibility of customizing the feature score function. Storing the data in alternative encoding better fit our objective of feature selection with high-dimensional datasets. While the limited amount of observations can be stored as columns without raising memory errors in the cluster, the high number of features can scale across the nodes. Hence, in the future, we intend to provide a portfolio of built-in feature selection algorithms that work with the alternative encoding. While we design and implement known FSA for MapReduce, novel algorithms that directly take advantage of the distributed nature of the data will be investigated as well. It might be interesting to extend the scalability study to classification tasks and network inference.

\ifCLASSOPTIONcaptionsoff
  \newpage
\fi



\bibliography{bibfile}
\bibliographystyle{IEEEtran} 
%



%

\begin{IEEEbiography}[{\includegraphics[width=1in,height=1.25in,clip,keepaspectratio]{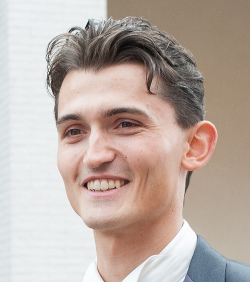}}]{Claudio Reggiani}
received his M.Eng degree in Computer Engineering from the Polytechnic University of Milan in 2013. He is currently a PhD candidate at the Machine Learning Group and the Interuniversity Institute of Bioinformatics in Brussels, (IB)2, Belgium. His research interests include machine learning, big data and bioinformatics.
\end{IEEEbiography}

\begin{IEEEbiography}[{\includegraphics[width=1in,height=1.25in,clip,keepaspectratio]{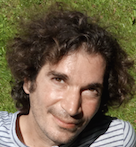}}]{Yann-A\"el Le Borgne}
is currently post-doctoral researcher at the Machine Learning Group and the Interuniversity Institute of Bioinformatics in Brussels (IB)2, Belgium. He received his Ph.D. degree in Computer Science from the University of Brussels, Belgium (Marie Curie fellowship), and MSc in Cognitive Sciences from the University Joseph Fourier, France. His research interests include machine learning, big data, bioinformatics, open data and reproducible research.
\end{IEEEbiography}


\begin{IEEEbiography}[{\includegraphics[width=1in,height=1.25in,clip,keepaspectratio]{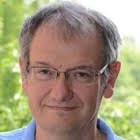}}]{Gianluca Bontempi}
is Full Professor in the Computer Science Department at the Universit\'{e} Libre de Bruxelles (ULB), Brussels, Belgium, co-head of the ULB Machine Learning Group and Director of (IB)2, the ULB/VUB Interuniversity Institute of Bioinformatics in Brussels. His main research interests are big data mining, machine learning, bioinformatics, causal inference, predictive modelling and their application to complex tasks in engineering (forecasting, fraud detection) and life science. He was Marie Curie fellow researcher, he was awarded in two international data analysis competitions and he took part to many research projects in collaboration with universities and private companies all over Europe. He is author of more than 200 scientific publications, associate editor of PLOS One, member of the scientific advisory board of Chist-ERA and IEEE Senior Member. He is also co-author of several open-source software packages for bioinformatics, data mining and prediction.
\end{IEEEbiography}




\end{document}